# Heterogeneous Freeform Metasurfaces: A Platform for Advanced Broadband Dispersion Engineering

Zhaoyi Li[1,2,†], Sawyer D. Campell[3,†], Joon-Suh Park[2,†], Ronald P. Jenkins[3], Soon Wei Daniel Lim[2], Douglas H. Werner[3], Federico Capasso[2]

1. Department of Materials Science and Engineering, Massachusetts Institute of Technology, Cambridge, MA, 02139, USA
2. Harvard John A. Paulson School of Engineering and Applied Sciences, Harvard University, Cambridge, MA, 02138, USA
3. School of Electrical Engineering and Computer Science, The Pennsylvania State University, University Park, PA, 16802, USA

[†]These authors contributed equally

**Abstract**

Metasurfaces, with their ability to control electromagnetic waves, hold immense potential in optical device design, especially for applications requiring precise control over dispersion. This work introduces an approach to dispersion engineering using heterogeneous freeform metasurfaces, which overcomes the limitations of conventional metasurfaces that often suffer from poor transmission, narrow bandwidth, and restricted polarization responses. By transitioning from single-layer, canonical meta-atoms to bilayer architectures with non-intuitive geometries, our design decouples intrinsic material properties (refractive index and group index), enabling independent engineering of phase and group delays as well as higher-order dispersion properties, while achieving high-efficiency under arbitrary polarization states. We implement a two-stage multi-objective optimization process to generate libraries of meta-atoms, which are then utilized for the rapid design of dispersion-engineered metasurfaces. Additionally, we present a bilayer metasurface stacking technique, paving the way for the realization of high-performance, dispersion-engineered optical devices. Our approach is validated through the demonstration of metasurfaces exhibiting superior chromatic aberration correction and broadband performance, with over 81% averaged efficiency across the 420-nm visible-to-near-infrared bandwidth. Our synergistic combination of advanced design physics, powerful freeform optimization methods, and bi-layer nanofabrication techniques represents a significant breakthrough compared to the state-of-the-art while opening new possibilities for broadband metasurface applications.

**Introduction**

Metasurfaces exhibit potential for revolutionizing conventional optical device design by offering unprecedented control over electromagnetic wave propagation[1-3]. This capability arises from the intelligent patterning subwavelength features, which are carefully designed to interact with incident wavefronts in a desired way. Designers often arrange these features in a regular, albeit occasionally aperiodic, fashion, selecting them from a library of 'meta-atoms'[4]. Unlike traditional bulk optical components, which are constrained by their comprising material's intrinsic chemical and atomic compositions, meta-atoms are artificial structures: Their optical properties—such as phase, dispersion, and polarization response—can be precisely tailored by manipulating their geometries. Moreover, metasurfaces can exhibit material properties that are not naturally found in conventional bulk materials. These properties include anomalous transmission/reflection and one-way transmission[5]. One of the most impactful features of metasurfaces lies in their ability to engineer dispersion behaviors[6-8]. In contrast to bulk materials, which have fixed dispersion characteristics, metasurfaces allow for nearly arbitrary control over dispersion. By patterning metasurfaces with judiciously designed meta-atoms, one can tailor their dispersion to specific needs. This control is crucial in optical designs, ensuring high-quality images by minimizing chromatic aberrations that can otherwise lead to smeared colors and blurry results. While strategies to correct for dispersion have been known for ages, they often involve pairing multiple distinct glasses together, resulting in bulkier and more expensive optical systems. In contrast, dispersion-engineered metasurfaces offer a compelling alternative. These metasurfaces achieve achromatic focusing in planar form factors, making them highly attractive for reducing system SWaP (size, weight, and power)[9,10]. Despite their promise, the dispersion engineering performance of current metasurface platforms is fundamentally constrained by the quality of available meta-atoms. Existing dispersion-engineered meta-atoms suffer from limitations such as poor transmission, narrow bandwidth, and restricted polarization responses. Additionally, while meta-atom geometries can be tuned to alter their phase responses, the coupling between phase and group delay (GD) constrains their dispersion engineering capabilities. Consequently, there is a critical need for meta-atoms that exhibit broadband, polarization-insensitive, high-transmission responses free from geometrical restrictions.

Numerous studies have highlighted the superiority of freeform (*i.e.*, unparameterized or unintuitive) metasurfaces over canonical designs, not only in terms of efficiency but also multifunctionality and robustness[11-13]. Freeform metasurface devices (meta-devices) excel at demonstrating various phenomena, including polarization splitting, multi-band beam steering, single photon emitters, and optical logic gates[14-17]. However, designing such devices lacks analytical solutions based on physics intuition; instead, numerical optimization is necessary[18]. The prevailing freeform optimization technique relies on the adjoint method (*i.e.*, topology optimization), which presents several challenges for realizing dispersion-engineered meta-devices. Firstly, while adjoint methods have been used to design entire lenses, their application becomes computationally intractable for devices on the order of a few tens of wavelengths due to the computational expense of large full-wave simulations. Secondly, adjoint methods are typically used to design supercells rather than meta-atom libraries due to restrictions in the merit function formulation. For example, maximizing reflection/transmission efficiency is straightforward, but achieving 2π phase coverage poses difficulties. Thirdly, topology optimization involves averaging gradient contributions throughout the volume to maintain material homogeneity in the vertical direction[19]. However, this process can negatively impact solution convergence and becomes less accurate for high-aspect-ratio structures. Furthermore, optimizing multiple objectives relies on a single weighted objective, requiring informed choices about balancing goals and constraints. Adjoint methods also suffer from local minima of the merit function in the design-response space, necessitating multiple trials with varying starting points. Complex fabrication constraints exacerbate this issue by introducing additional gradients[20] that may compete with device performance during convergence to a binary structure. In contrast, our approach generates entire libraries of meta-atoms. These meta-atoms have their transmission/phase/dispersion hypervolumes[21] maximized, offering a wide range of optimized building blocks. This strategy enables designers to rapidly synthesize highly performant meta-devices while adhering to fabrication constraints. Additionally, it provides flexibility in realizing large area designs with varied dispersion engineered performances. Nevertheless, the performance upper bound of any metasurface is ultimately limited by the underlying physics, regardless of the optimization tool's power.

A single-layer metasurface, comprising meta-atoms of simple geometry, inherently suffers from chromatic aberrations (Fig. 1(a)). Upon normal incidence of a light pulse, the metasurface diffracts various colors at different angles, adhering to the generalized Snell's law[22,23]. The phase delay within such a device is a function of the meta-atom's geometry, influencing its effective index of refraction. Unfortunately, the dispersion properties, such as group delay, are not independently tunable. This limitation constricts the operation bandwidth, though the efficiency can be high. To address the dispersion challenges in metasurfaces for broadband applications, complex meta-atoms have been utilized[24-27]. A notable example is the double-fin meta-atom (Fig. 1(b)), which, with its rotational freedom, facilitates dispersion control[24]. The group index of such a meta-atom is engineered by its geometry, which governs light confinement within the nanostructure. It can be conceptually understood by picturing meta-atoms as waveguides[28,29]. Rotating the meta-atom by an angle of α imparts a Pancharatnam-Berry (PB) phase of 2α, achieved via polarization conversion[30,31]. Nonetheless, such metasurface operation imposes polarization limitations, stemming from the PB phase's physical origins. As depicted in Fig. 1(b), a dispersion-engineered metasurface can achromatically redirect light from a left-handed to a right-handed polarization state. However, the primary drawback is the typically low polarization conversion efficiency, which further diminishes when straying from the central design wavelength[24,25].

To address the current limitations of dispersion engineering, which often result in constrained polarization performance and low efficiency, we pursued two synergistic pathways. First, we elevated the structural dimensionality by transitioning from a single-layer to a bilayer architecture. Second, we expanded the design space dimensionality by transitioning from simple meta-atoms with canonical shapes to freeform meta-atoms featuring non-intuitive geometries. By combining both structural and design space scaling, we heterogeneously stacked two layers, creating surrealistic building blocks for bilayer freeform metasurfaces (Fig. 1(c)). These bilayer structures not only decouple the intrinsic material properties of effective index and group index (which determine phase and group delays, respectively), but also allow independent engineering of higher-order group delay dispersion behaviors. Furthermore, these heterogeneously stacked bilayer freeform meta-atoms operate under arbitrary polarization states with extremely high efficiency. Our process leverages state-of-the-art design, optimization, and nanofabrication techniques, unlocking the new physics of dispersion-engineered meta-atoms and enabling rapid synthesis of a wide range of dispersion-engineered meta-devices.

To optimize the design of freeform meta-atoms, we implement a two-stage multi-objective optimization (MOO) process[32], as summarized in Fig. 2. The flowchart (see Fig. 2(a)) initiates with the exploration of the fundamental "design-physics", aiming to decouple two fundamental material properties: refractive index and group index. The refractive index describes how much light slows down when entering a medium compared to its speed in a vacuum. On the other hand, the group index quantifies the reduction in group velocity, the speed of a light pulse, when light propagates within a medium[33]. Additionally, the group index reflects the dispersion characteristics of the material. In bulk materials, these properties are typically fixed and coupled. However, in nanostructured materials with subwavelength feature sizes, we can engineer the refractive index and group index independently. In metasurfaces, the effective refractive index and group index depend not only on material compositions but also on their constituent nanostructures' geometries[15,34-36]. For example, controlling the filling ratios of one material in its host matrix within a meta-atom allows us to alter its effective refractive index. Figure 2(b), the blue and red lines, illustrate the relationship between phase delay and group delay in a square nanofin with varying material compositions as its fin size changes. Generally, increasing the filling ratio of a high-index material leads to a higher effective index. However, in simple canonical meta-atoms, phase delay and group delay are correlated due to limited design freedoms (Fig. S6). To decouple these two quantities, one can introduce more design parameters and engineer complex-shaped meta-atoms, and thus the optical mode confinement inside the nanostructure. For example, a rectangular nanofin with two design parameters can decouple phase delay from group delay within a limited space (Fig. S7). A double-fin meta-atom possessing five design parameters allows larger decoupling range. For even greater flexibility, meta-atoms with freeform shapes (blue dashed and red dashed regions in Fig. 2(b)) are desirable. However, the extent of decoupling achievable through this method remains bounded by constituent material compositions; In many cases, the phase coverage falls short of 2π for a given group delay, limiting the performance of dispersion-engineered metasurfaces. To overcome this limitation, we propose heterogeneously stacking two meta-atoms with distinct material compositions. By doing so, we significantly expand the decoupling space. For example, stacking of two simple meta-atoms (*e.g.*, square nanofins) can significantly broaden the decoupling space to the grey area outlined by solid black lines. In a first-order approximation, the phase delay and group delay of a bilayer meta-atom can be treated as the sum of their individual layer contributions (Fig. S8). On top of that, by carefully shaping meta-atoms in a freeform manner, we can achieve an even larger decoupling range (grey area outlined by dashed black lines), enabling the custom design of superior dispersion-engineered metasurfaces.

In the subsequent step, we generate an initial set of design vectors. Each design vector is input into the geometry synthesis algorithm. For bi-layer designs, this algorithm controls the inclusions in the bottom layer and shapes the freeform top-layer meta-atom. The techniques introduced in reference[4] guide the manipulation of control points, which govern the shape of a Bézier (or spline) surface. This continuous surface is then binarized using a level-set procedure to realize the meta-atom mask. The generated masks undergo a fabrication enforcement process based on an intelligent use of morphological "open" and "close" operations[37]. This process ensures minimum feature size widths and areas, as well as gaps between meta-atoms within a single unit cell. The resulting meta-atoms still maintain their freeform nature, but with enforced fabrication constraints (see Fig. 2(c)). To evaluate their performance, we simulate the candidate meta-atoms using Rigorous Coupled Wave Analysis (RCWA). This computational approach is particularly efficient for all-dielectric meta-atoms[38]. The simulations cover a bandwidth from 480 nm to 680 nm. We average the transmission magnitudes and fit the frequency-dependent phase response with a second-order Taylor series expansion. This allows us to capture the phase delay at the expansion angular frequency $\omega_d$ (560nm) as well as the group delay (GD) and group delay dispersion (GDD) coefficients. Next, we define four optimization objectives using a coordinate transformation process described in reference[4]. Our goal is to simultaneously maximize the phase, GD, and GDD ranges. Unlike conventional optimization approaches that seek to find one design with a targeted complex transmission characteristic, our implemented MOO process is much more efficient and generates an entire library of optimized meta-atoms in a single "click." The results of multiple

single- and bi-layer optimizations, incorporating 2- and 4-fold symmetries, are shown in Fig. 2(d). The optimized meta-atoms are able to span a wide range of phase, GD, and GDD values within the library. This is highlighted in Fig. 2(e) where five meta-atoms are shown to possess the same GD, but with varying phase delays and in Fig. 2(f) where another set of five meta-atoms possess the same phase delay, but with varying GD values. This empowers the designer to realize a wide range of dispersion-engineered metalens solutions.

The second stage of the MOO process begins once the designer has determined the spatial distributions of phase delay, GD, and GDD values required to achieve a metasurface with a desired dispersion-engineering performance. During this stage, the optimizer aims to identify a subset of designs from the library that simultaneously minimizes phase, GD, and GDD errors across the surface while maximizing transmission. The optimal compromise solution is then selected from the Pareto set. For example, in Fig. 2(g), we illustrate the reconstructed phase profiles and dispersion profiles of an optimized achromatic metalens, in comparison to their target values. This selected design is then used to generate GDS-II files, which contain the arrangement of meta-atoms patterned across the top and bottom layers. Finally, the resulting optimized metalens can be validated through full-wave simulations, such as Finite Difference Time Domain (FDTD), to assess its design performance.

Figure 3(a) illustrates a cross-section view of a bilayer meta-atom. In general, the nanostructures in both the top and bottom layers possess anisotropic shapes and mirror symmetries about the XZ or YZ planes. When light propagates within these meta-atoms, various polarization-dependent modes emerge in each layer. Figures 3(b) and (c) are the simulation results of electric field distribution in the top layer (top view of Fig. 3(a)) under x- and y-polarized incidences, respectively. The incident wavelength is 535 nm, and light propagates from the bottom to the top. Under x-polarized incidence, the electric field energy primarily concentrates in the air gap. Conversely, under y-polarized incidence, electric field confinement occurs mainly in the two arms of the meta-atom, where the main body consists of high-index titanium dioxide ($TiO_2$). Figures 3(e) and (d) reveal distinct electric field confinement in the bottom layer under x- and y-polarized incidences, respectively. One elongates along the x-direction, while the other elongates along the y-direction. The meta-atoms are carefully designed to ensure that x- and y-polarized light, as it propagates through the meta-atoms, accumulates a similar phase delay at the end. This behavior is evident from the simulation results of the phase distribution in the XZ plane (Fig. 3(d)) and YZ plane (Fig. 3(g)) when the incident polarization is along x and y, respectively. Despite experiencing distinct perturbation patterns while traveling through both layers, as indicated by the fine features in the phase map, the final accumulated phases in the air remain similar. Moreover, this holds valid across a broad bandwidth (Fig. S10). This key feature enables broadband polarization-insensitive performance and is achieved by optimizing the combinations of anisotropic structures in stacked heterogeneous layers. Our design strategy offers a significant advantage over conventional polarization-insensitive approaches that enforce 4-fold symmetry onto the meta-atoms. By allowing more diverse topology, our design provides a greater decoupling regime in the dispersion diagram (Fig. 2(d)). Notably, our heterogeneous meta-atoms also include designs with 4-fold symmetry, further enhancing dispersion coverage.

One of the primary reasons that freeform and bilayer metasurfaces for visible wavelengths remain unexplored is the substantial fabrication challenges they present. Previous research efforts to create bilayer metasurfaces often involved bonding two separately fabricated metasurfaces[39] or cascading two metasurfaces with an air gap[40]. These approaches require complex alignment hardware and are prone to unwanted interlayer scattering and efficiency loss. In other efforts, a more direct fabrication method was pursued: a first layer metasurface was fabricated on a substrate, followed by coating the surface with an optically clear dielectric such as a polymer, and then fabricating another layer on the polymer[41,42]. This approach allows creating a bilayer metasurface on a single substrate. However, the use of a polymer encapsulation layer complicates the surface planarization process and may lead to degradation or swelling when exposed to a solvent-based process. Additionally, uniform, conformal coating of polymers on dense and high-aspect-ratio nanostructures is difficult to achieve due to surface tension and wetting issues. To address these challenges, we adopt a more direct fabrication approach, as summarized in Figure 4(a).

The fabrication process begins with defining the bottom embedded meta-atoms by etching high-aspect-ratio trenches into glass and conformally filling them with $TiO_2$ dielectrics with an atomic layer deposition process. The surface of the bottom layer is then planarized by plasma etching, which can also be done with chemical mechanical polishing. The top layer meta-atoms are then directly fabricated on this prepared base using any fabrication process that ensures precise alignment. Our method ensures that the two metasurface layers are in direct contact, minimizing inter-layer scattering and the unwanted interactions between a meta-atom in a different layer. Thus, this platform allows us to achieve high efficiency at the meta-atom level.

In this study, we employ an aligned electron-beam lithography (EBL) method to define top layer meta-atoms with freeform shapes directly into the e-beam resist film. Details of the fabrication flow are provided in the Method section. While high-aspect-ratio and small-feature-size freeform meta-atoms are ideal for visible metasurfaces, they are difficult to realize with a pattern transfer and etch process. The accuracy of realizing such structures is prone to fabrication errors such as an imperfect etch mask formation or the etch sidewall profile. Our fabrication method avoids direct etching of freeform meta-atoms which allows formation of high-aspect-ratio, small features in a high spatial density, and our designs are optimized with the resolution limit and the EBL accuracy taken into consideration. The design and fabrication flow work synergistically to accurately define freeform meta-atoms, resulting in a 600-nm-height freeform $TiO_2$ meta-atom atop a 1500-nm-thick $TiO_2$-in-$SiO_2$ embedded nanocomposite meta-atom. Examples of the top-layer freeform meta-atoms on patterned substrate are highlighted in a tilted-view scanning electron microscope image (Fig. 4(b)). Additionally, Fig. 4(c) shows a cross-sectional image of the bilayer meta-atoms obtained through focused-ion beam milling, revealing both the top and the bottom nanostructures.

Precise control of meta-atom dispersion is crucial for achieving high-efficiency broadband metasurfaces. As an illustrative example, Fig. 5 presents measurement results for a chromatic metalens composed of dispersion-engineered freeform meta-atoms. The metalens has a diameter of 110 microns, as revealed in the scanning electron microscope (SEM) image (Fig. 5(a)). Its designed numerical aperture at a wavelength of 560 nm is 0.1. Zoomed-in SEM images from titled and top views are shown in Figs. 5(b) and 5(c), respectively. The metalens consists of a single layer of freeform titanium dioxide meta-atoms. Figure 5(d) provides a perspective view, highlighting densely packed meta-atoms with a diverse range of morphologies along the metalens edge. These meta-atoms are intentionally designed with two-fold symmetry to facilitate polarization engineering. The chromatic metalens implements phase profiles as:

$$\varphi(\omega, r) = \frac{\omega}{C}\left(f(\omega) - \sqrt{f(\omega)^2 + r^2}\right) \text{ (Eq.1)},$$

where $\omega$ is the angular frequency of light, $C$ is the speed of light in a vacuum, $f(\omega)$ is the focal length, and $r$ is the radial coordinate.

The focal length follows a chromatic dispersion:

$$f(\omega) = f(\omega_0) \times \frac{\omega}{\omega_0} \text{ (Eq.2)},$$

where $\omega_0$ is the center angular frequency.

By applying Taylor's expansion to Eq.1, one can deduce the required phase profile at the center design wavelength, the required group delay and group delay dispersion as:

$$\varphi(\omega_0, r) = \frac{\omega_0}{C}\left(f(\omega_0) - \sqrt{f(\omega_0)^2 + r^2}\right) \text{ (Eq.3)},$$

$$GD(r) = \left.\frac{\partial \varphi(\omega, r)}{\partial \omega}\right|_{\omega_0} \text{ (Eq.4)},$$

where $GD$ denotes group delay;

$$GDD(r) = \left.\frac{\partial^2 \varphi(\omega, r)}{\partial \omega^2}\right|_{\omega_0} \text{ (Eq.5)},$$

where $GDD$ denotes group delay dispersion. The designed phase profiles and dispersion profiles, compared to their implemented values, of the freeform meta-atoms are summarized in the supplementary information. One can tell that the chromatic metalens needs mostly meta-atoms of similar dispersion behavior. Figures 5(e)-(j) summarize the measurement results. Figure 5(e) is the measured point spread functions in the XZ cross section, where X corresponds to the metalens radial direction. The measurements were conducted in the visible band, spanning wavelengths ranging from 480 nm to 680 nm. The measured focal shift, $\Delta F$, is 212 microns, which corresponds to a relative focal shift $\Delta F/F(\lambda_{center})$ of 41.4% with respect to the center wavelength, $\lambda_{center}$, of 580 nm. For comparison, FDTD simulation results (the SI) show a simulated focal shift of 200 microns and a relative focal shift of 38.4%. To demonstrate diffraction-limited focusing performance, Fig. 5(f) shows the measured focal points at each focal plane and at wavelengths of 480 nm, 580 nm, and 680 nm. Additional measurements at other wavelengths are detailed in SI. Figure 5(g) shows the measured point spread functions (XZ cross section) in the near infrared band ranging from 700 nm to 900 nm in wavelength. The measured focal shift is 104 microns, and relative focal shift is 28.4%. In comparison, the FDTD simulation (Fig. S13) predicts a focal shift of 96 microns and relative focal shift of 25.7%. The results in Fig. 5(h) (more in SI) confirm diffraction-limited focusing in the NIR. The measured Strehl ratio (Fig. 5(i)) ranges from 0.93 to 0.97 across the VIS-NIR bandwidth, which exceeds the 0.8 criteria. Averaged focusing efficiencies are 84.8% in the visible band, 92.2% in the NIR band, and 88.5% in the whole 420 nm bandwidth. Corresponding simulation values are 92.9%, 97.2%, and 95.2%, respectively. Their comparison is shown in Fig. 5(j).

To demonstrate the significance of our dispersion-engineered method for a metasurface demanding dispersion-diverse meta-atoms, we further designed and fabricated an achromatic metalens using the heterogeneous freeform meta-atom platform. The design of the achromatic metalens enforces a constant focal length for all wavelengths:

$f(\omega) = f_o$ (Eq.6).

This implies that the required group delay profile is

$GD(r) = \frac{1}{c}\left(f_o - \sqrt{f_o^2 + r^2}\right)$ (Eq.7),

and the required group delay dispersion profile is constant:

$GDD(r) = 0$ (Eq.8).

To achieve this, we employed a multi-objective inverse-design process to simultaneously optimize the phase and dispersion profiles. The implemented phase profiles, obtained through judicious selection of meta-atoms from the library, closely match their target values at visible wavelengths, as shown in SI. Figures 6(a)-(f) showcase the fabrication results of the achromatic metalens. It was designed with a diameter of 110 microns and a numerical aperture is 0.1 for direct comparison to the previously discussed chromatic metalens. Figures 6(b) and (c) are the top-view SEM images of the center and the edge of the device. The device comprises three distinct types of meta-atoms: namely, a top-layer freeform meta-atom (similar to the ones used in the chromatic metalens), a bottom-layer meta-atom by embedding $TiO_2$ nanostructures inside fused silica, and a bilayer freeform meta-atom formed by stacking the top and bottom layers. The different meta-atom regions are highlighted in Fig. 6(c). For example, in the region consisting only of bottom-layer meta-atoms, one sees empty voids on the top surface of the device, exposing the patterned substrate. Zooming in, Figs. 6(d)-(f) provide detailed views of the meta-atoms from a tilted angle. The top-layer meta-atom has a height of 600 nm, while the bottom-embedded meta-atom measures 1500 nm in thickness, resulting in a total thickness of 2100 nm. Figure 6(f) highlights the transition from the bi-layer region to the bottom-layer region. The minimum feature size is 50 nm in the top layer and 80 nm in the bottom layer, indicating an aspect ratio greater than 30 considering the stacked layers.

Figure 7 summarizes the measurement results of the fabricated broadband achromatic metalens. In Fig. 7(a), we show the measured cross sections of the point spread functions in the XZ planes across wavelengths from 480 nm to 680 nm. The maximum focal shift measured within the wavelength range is 28 microns. Considering the measured focal length at the center wavelength of 580 nm to be 530 microns, the relative focal shift is as low as 5.28%. The FDTD simulation results are shown in the SI, where the simulated maximum focal shift is 18 microns, and relative focal shift is 3.4%. Figure 7(b) shows the measured results in the NIR band (700 nm to 900 nm). Here, the measured maximum focal shift is 45 microns, and the focal length at the center wavelength of 800 nm is measured to be 462 microns. It results in a relative focal shift of 9.74%. The simulation (Fig. S15) predicts a maximum focal shift of 60 microns and a relative focal shift of 12.2%. By comparing the chromatic metalens and the achromatic metalens side by side, we showcase the dispersion-engineering results (SI). The measured chromatic dispersion, quantified by relative focal shift, changes from 41.1% to 5.2% in the visible, corresponding to a reduction of 87.3% in chromaticity. In the NIR band, the relative focal shift decreases from 28.4% to 9.7%, corresponding to a 65.8% reduction in chromaticity. The FDTD simulation results show a dispersion reduction of 91.1% in the visible and 52.9% in the NIR. The slight discrepancies between measurements and simulations may be attributed to the fabrication errors in bottom layer meta-atom thickness, leading to a red shift in the center operation wavelength. Figure 7(c) shows examples of the measured normalized intensity distributions of the focal spots at different wavelengths. Additional measurement results can be found in SI. In Fig. 7(e), we show the obtained Strehl ratio, which ranges from 0.92 to 0.98, indicating diffraction-limited performance across the VIS-NIR bandwidth. Comparing the measured focusing efficiencies with simulation results (Fig. 7(f)), the measured focusing efficiency is 74.7% in the visible band, 87.4% in the NIR band, and 81.0% across the whole measured spectrum. Corresponding simulation values are 83.6%, 92.0%, and 87.8%. The measured focusing efficiency gradually increases from short to long wavelengths. This behavior suggests that the scattering loss of meta-atoms, more pronounced at shorter wavelengths, plays a key role in determining efficiency. Achieving high efficiency requires deep subwavelength meta-atom sizes across the entire operation wavelength range. However, this imposes challenges in fabrication, especially when extending the working bandwidth into the visible regime.  We also compared the focusing efficiencies of the chromatic metalens and achromatic metalens side by side in the SI. Existing dispersion engineering techniques often come with a substantial efficiency loss. Our findings demonstrate that using a heterogeneous and freeform platform for dispersion engineering can achieve an average loss of as low as 7.4%, suggesting a paradigm shift towards high-performance dispersion-engineering methodologies.

**Discussion**

Dispersion engineering concerns a long-standing fundamental question in physics. In the context of conventional refractive optics, dispersion stems from intrinsic material properties[43]. Traditionally, addressing dispersion involves combining multiple optical elements made from diverse materials. Yet, this approach faces limitations: First, the Abbe number, a key metric characterizing material dispersion, is positive for homogeneous materials[44]. Additionally, the material dispersion is constrained to discrete values according to the refractive index-Abbe number diagram[45], and its fine-tuning is challenging[46]. These lead to a trade-off between dispersion engineering performance and design compactness. Second, high-performance dispersion engineering often requires materials with exotic dispersion properties. Unfortunately, these materials tend to be expensive and pose significant manufacturing challenges due to their mechanical properties. Third, achieving precise, spatially varying dispersion engineering remains elusive within this framework.

In our study, we introduced a paradigm shift in the realm of dispersion engineering by employing heterogeneous freeform metasurfaces. Unlike conventional homogeneous material stacking methods, we not only leveraged spatially varying heterogeneous material compositions, but also exploited the rich optical modes confined within freeform nanostructures. By meticulously controlling light propagation across orthogonal polarization states, we achieved polarization-insensitive performance through the optimization and stacking of anisotropic nanostructures. These demonstrated heterogeneous freeform meta-atoms not only cover a wide dispersion range but also exhibit high transmission efficiency across a broad bandwidth. To design these non-intuitive meta-atoms, we presented a two-stage multiple-objective optimization framework that sequentially optimizes the meta-atom library and meta-device. Furthermore, we introduced a metasurface stacking technique for precise realization of bilayer freeform metasurfaces, with potential extensions to multi-layered (≥ three layers) metasurfaces in the future.

Our current design accounts for interlayer coupling by jointly modeling both layers. However, the intralayer coupling is neglected by using the local periodic approximation[47]. For metasurfaces with abrupt phase or dispersion changes, intralayer coupling may impact performance. Recent advancements in specialized finite element methods (FEM) and deep learning seek to accelerate modeling and prediction of coupling between elements[48]. Integrating these capabilities into our proposed design methodology can not only improve agreements between simulated and fabricated devices, but also lead to better performing designs. Fabricating freeform meta-atoms, particularly those with small features and high aspect ratios, presents significant challenges. For instance, the side walls of our bottom-layer embedded meta-atoms are not perfectly vertical, necessitating improvements in etching techniques. Our demonstrated dispersion-engineered meta-device has a size approximately 200 times the operation wavelength. To further increase the device size, we could utilize a multi-layer metasurface platform or a hybrid platform that integrates metasurfaces with refractive optics. Such a hybrid platform can relax the requisite dispersion range needed for the metasurface to correct for color while exploiting the refractive element which provides the bulk of the focusing power. Additionally, due to the increased number of parameters inherent in freeform meta-atoms compared to simple-shaped ones, the resulting GDSII files for freeform metasurfaces tend to be quite large. Addressing the design file compression without sacrificing key features is a crucial consideration for realizing large-area freeform metasurfaces in the future

Our platform, which utilizes heterogeneous freeform metasurfaces, offers a new approach to high-performance dispersion engineering. This technology has the potential to unveil a multitude of intriguing physical phenomena. For example, it can facilitate a high-dimensional, multifunctional meta-device that exhibits spin-, polarization-, and angular-dependent responses. These optical responses are crucial for optical encoders, which play a vital role in optics-based artificial intelligence systems for multi-channel information processing[49-51]. Moreover, our platform can synergize with other optical technologies[52]. By integrating freeform metasurfaces with refractive counterparts, one can achieve high-performance, compact, broadband hybrid optics with large aperture sizes[53]. We envision our platform paving the way for exciting applications in diverse fields, including virtual reality, augmented reality, optical computing, machine vision, and sensing.

**Method**

Fabrication

The fabrication process is summarized as follows. First, the bottom layer of the meta-atom is fabricated using a 150 nm thick aluminum (Al)-coated fused silica wafer, which is spin-coated with an electron-beam (e-beam) resist (950 PMMA, Kayaku Advanced Materials). After e-beam lithography (EBL) and development of the e-beam resist, the lithographed resist pattern is then transferred into the Al layer using $Cl_2$/Ar plasma etch. The resist layer is then stripped using downstream oxygen plasma etch, and the Al-defined pattern serves as an etch mask to etch the underlying fused silica wafer to a depth of 1.5 μm using fluoride plasma etch[54]. Note that the bottom layer meta-atoms are defined as trenches to minimize etch depth variance among different sizes due to the microloading effect[55] during the fused silica plasma etch. Subsequently, the Al layer is removed using a wet-etching process with 2.4 % tetramethylammonium hydroxide (TMAH) solution, leaving only the designed fused-silica trenches in the wafer. The wafer is then conformally coated with $TiO_2$ using atomic layer deposition (ALD) until the trenches are completely filled. The over-deposited $TiO_2$ layer is then etched away using $Cl_2$/Ar/$O_2$ plasma etch, completing the fabrication of the bottom $TiO_2$/$SiO_2$ embedded meta-atom layer. As fused silica is chemically protected against etching by $Cl_2$ plasma[56], the etching of the $TiO_2$ film naturally slows down once the $SiO_2$ layer is exposed, allowing better control over the etching process.

The top layer of the meta-atom is then fabricated on top of the as-prepared bottom layer. An e-beam resist (ZEP520A, ZEON SMI) is spin-coated onto the bottom layer to a thickness of 600 nm, which determines the thickness of the meta-atom's top layer[57]. Using an aligned EBL write, the top layer structures are defined as shaped holes in the resist film. These holes are then conformally coated with $TiO_2$ using a low-temperature ALD process until they are completely filled, resulting in a planarized surface. The excess $TiO_2$ layer is then etched away using $CHF_3$/Ar/$O_2$ plasma etch, until the resist layer is exposed. Finally, the resist layer is selectively etched away using downstream oxygen plasma etch, leaving only the free-form meta-atom's top layer structure.

**Acknowledgements**

Z. Li, J.P., S.W.D.L, and F.C. were supported by the Defense Advanced Research Projects Agency (DARPA) Grant No. HR00111810001. F.C., J.P., and S.W.D.L. were also supported by the Air Force Office of Scientific Research under award Number FA9550-21-1-0312. S.C., R.J. and D.W. were supported in part by the DARPA (HR00111720032). This work was performed in part at the Harvard University Center for Nanoscale Systems (CNS); a member of the National Nanotechnology Coordinated Infrastructure Network (NNCI), which is supported by the National Science Foundation under NSF award no. ECCS-2025158.

**Author contributions**

Z.L. and S.C. conceived the project. S.C., Z.L., and R.J. executed the designs and simulations. J.P. and Z.L. fabricated the devices, and Z.L. and J.P. performed the measurements. J.P. and Z.L. captured the scanning electron microscope images, while S.W.D.L. and J.P. conducted the focused ion beam characterization. D.W. and F.C. supervised the project. Z.L. and S.C. analyzed the data and wrote the first draft of the manuscript. All authors contributed to commenting on and editing the manuscript.

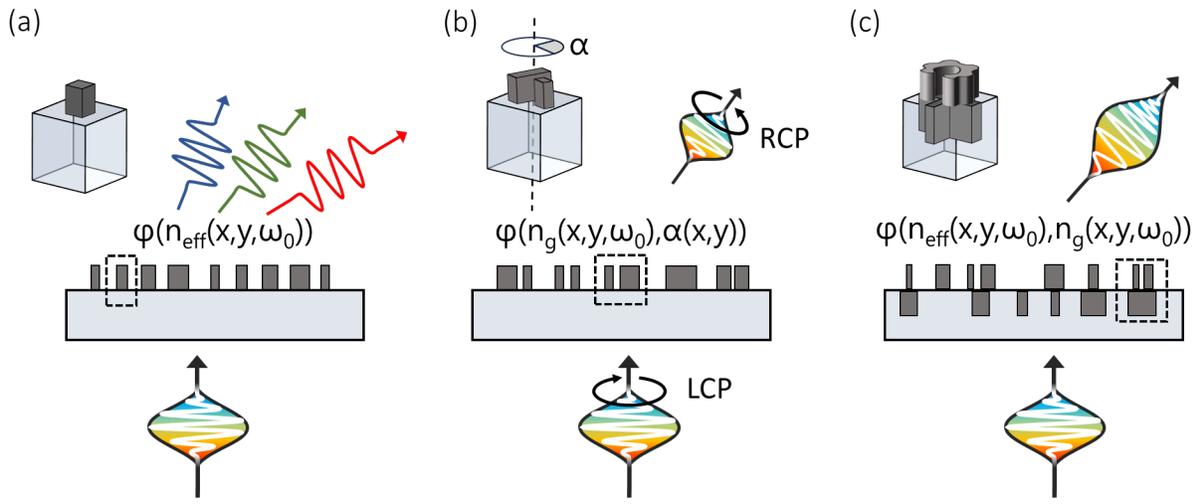

Figure 1. Comparison of three metasurface platforms. (a) A single-layer metasurface consisting of meta-atoms with simple geometries. An example of a meta-atom in the shape of square pillar is illustrated. (b) A single layer metasurface consisting of rotated meta-atoms with varied geometries. An example of double-fin meta-atom with a rotation angle of α is shown. (c) A bilayer metasurface by heterogenous stacking of freeform meta-atoms. The unit cell schematic shows the stacking of a freeform meta-atom on top of a cross-shape meta-atom that is embedded in substrate. The table below compares their performance in four aspects: dispersion engineering capability, operation bandwidth, polarization constraint, and efficiency.

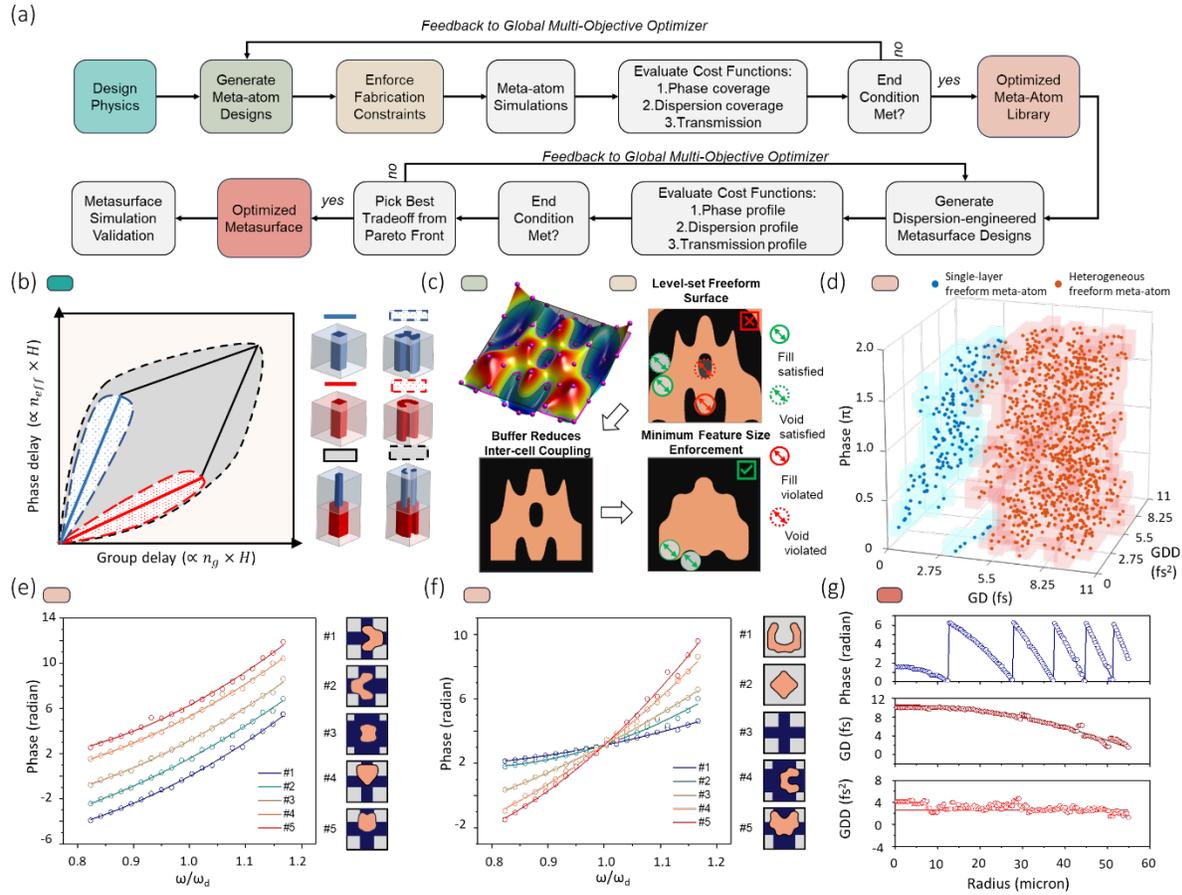

Figure 2. Design and physics of heterogenous metasurfaces. (a) The design flow contains two main optimization processes: the optimization of freeform meta-atom library and the optimization of metasurface based on the generated library. (b) Schematic of the design physics of the freeform heterogenous meta-atoms. They can decouple two fundamental material properties, refractive index and group index over an extended regime. (c) Generation of freeform meta-atoms by using the method of control points. The fabrication constraints are applied to enforce the minimum feature size. (d) The generated meta-atom library in the three-dimensional design space of phase, group delay, and group delay dispersion. The blue dot denotes single-layer freeform meta-atoms showing limited coverage in dispersion space. The red dot denotes heterogenous freeform meta-atoms that cover a much larger dispersion space. They can decouple the three physical terms. (e) A set of five meta-atoms from the library exhibits similar group delay but varied phase from 0 to 2π. (f) A set of five meta-atoms shows similar phase at the design wavelength but varied group delay. (g) The phase profiles of an achromatic metalens from 480 nm to 680 nm in wavelength. Solid lines represent the design values, and dots are the simulation values of meta-atoms.

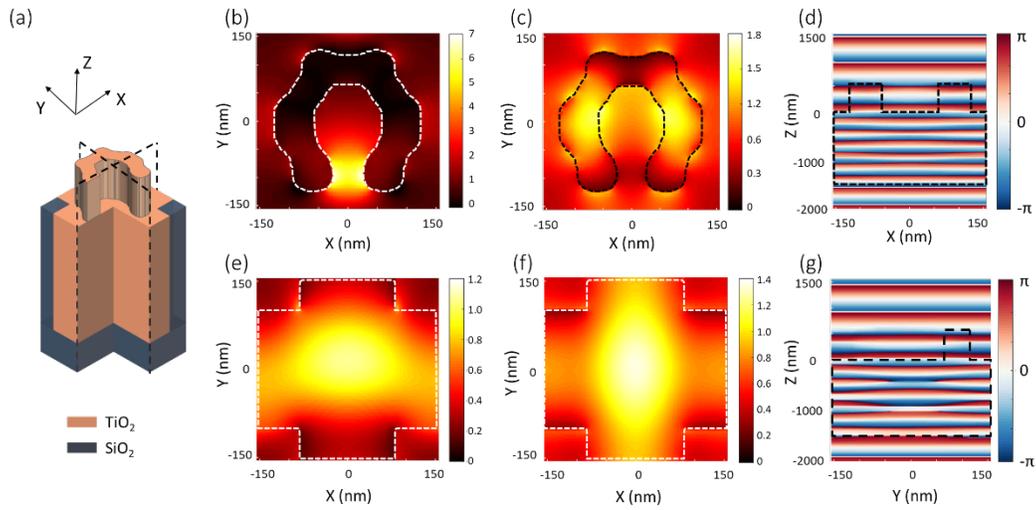

Figure 3. Polarization-insensitive performance of anisotropic heterogeneous meta-atoms. (a) Schematic of an anisotropic heterogeneous meta-atom. The material is coded in colors. (b) and (c) FDTD simulation results of electric field (absolute value) distribution in the XY plane of the top layer under X- and Y-polarized incident light. The simulation wavelength is 535 nm. (d) The phase map of the Ex-component of the field in the XZ cross section. Light propagates from the bottom to the top. (e) and (f) FDTD simulation results of the electric field distribution in the XY plane of the bottom layer under X- and Y-polarized incident light. (g) The phase map of the Ey component of the field in the YZ cross section.

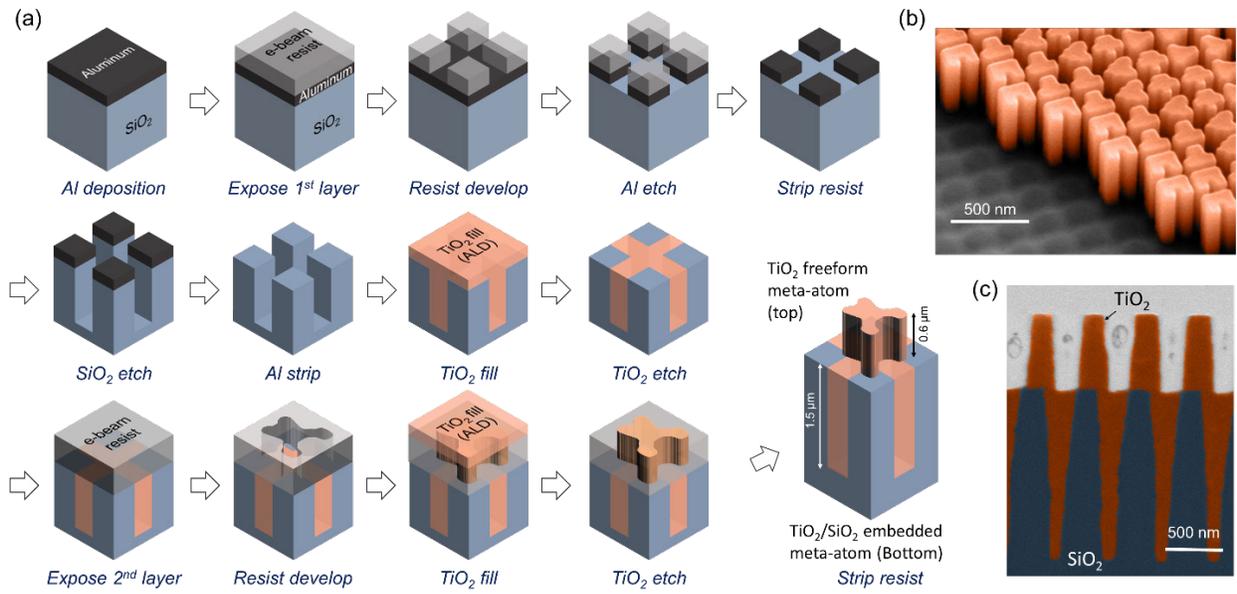

Figure 4. Fabrication of heterogeneous metasurfaces. (a) Schematic of fabrication flow using a unit cell as illustration. (b) Scanning electron microscopy (SEM) image of a fabricated device. The scale bar is 500 nm. (c) The SEM image of the cross section of the metasurface that is defined by using focused ion beam shows the stacking of two layers of meta-atoms. The scale bar is 500 nm.

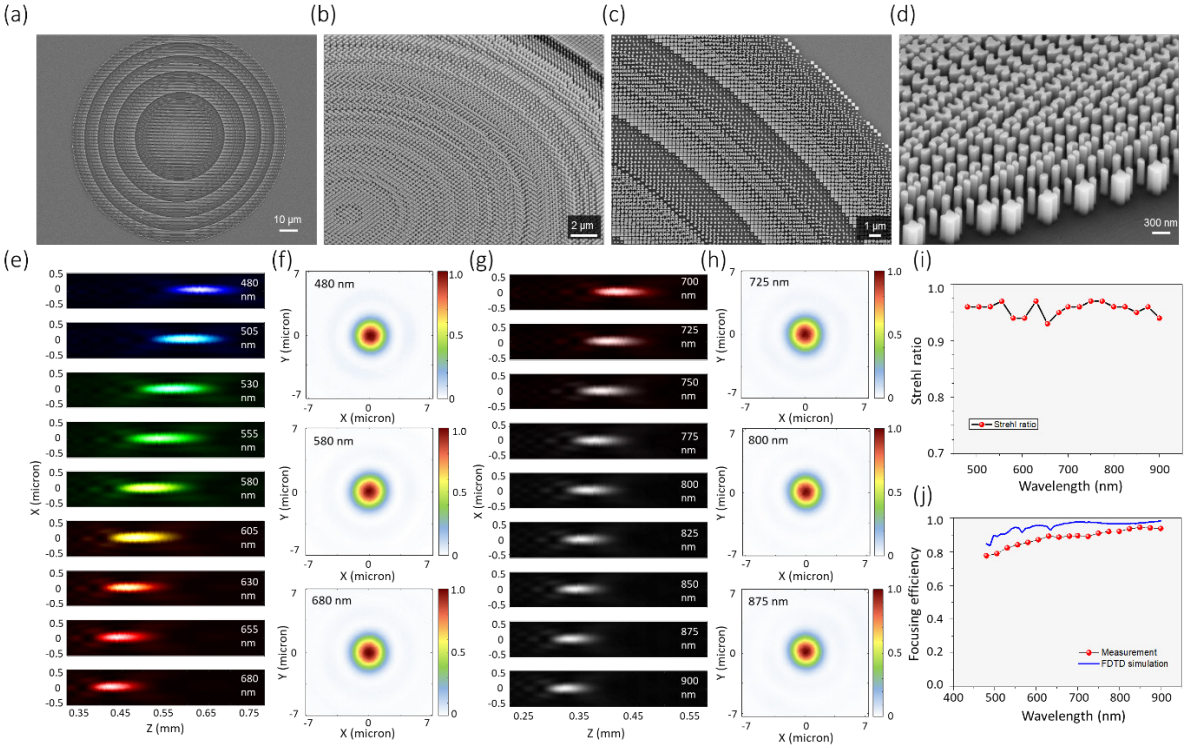

Figure 5. Measurement of a high-efficiency chromatic single-layer metalens. (a)-(d) SEM images of a fabricated single-layer metalens consisting of freeform meta-atoms. (a) and (c) are top views. (b) and (d) are titled views. (e) Measured point spread functions in the XZ plane, where Z is along the optical axis. The wavelength range is from 480 nm to 680 nm in the visible band. (f) Measured normalized light intensity distributions in the focal planes at wavelengths of 480 nm, 580 nm, and 680 nm, respectively. (g) Measured point spread functions in the XZ plane at wavelength ranging from 700 nm to 900 nm in the near infrared (NIR) band. (h) Measured normalized light intensity distributions in the focal planes at wavelengths of 725 nm, 800 nm, and 875 nm, respectively. (i) Measured Strehl ratio of the focal spots in the visible-NIR regime. (j) The measured focusing efficiency in the visible-NIR regime in comparison with simulation results.

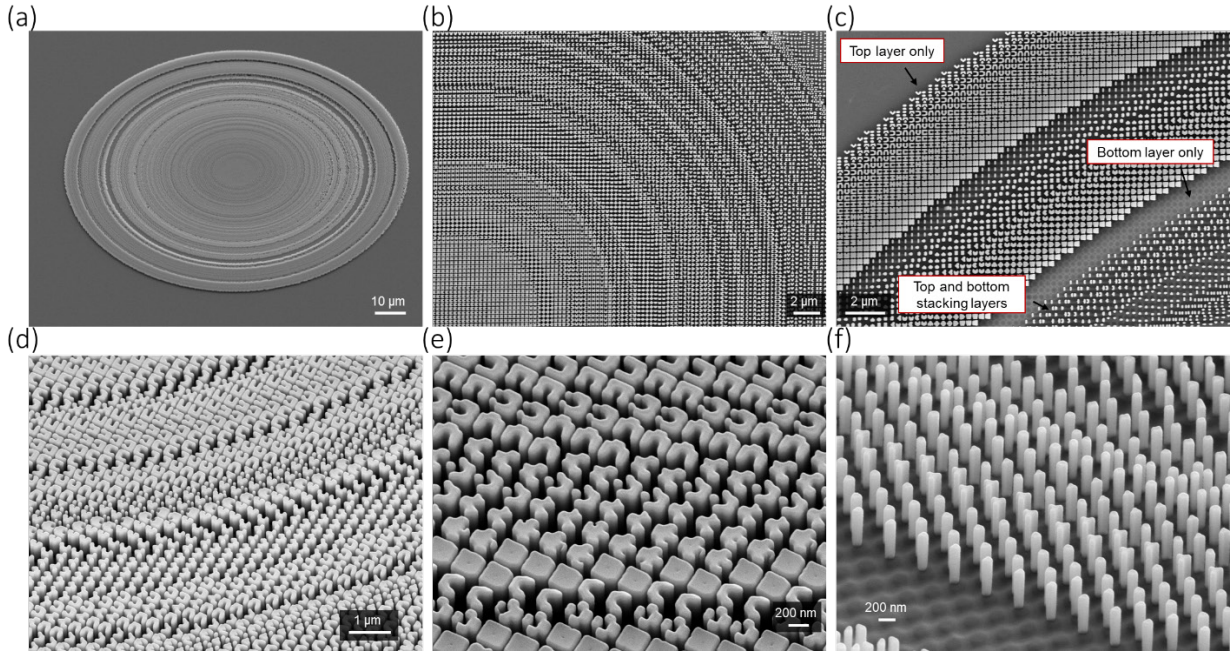

Figure 6. Fabrication of a high-efficiency dispersion-engineered heterogeneous metalens. (a) SEM image of a whole fabricated device. This device consists of three types of meta-atoms: top single-layer freeform meta-atoms, bottom single-layer $TiO_2$-embed-in-$SiO_2$ meta-atoms, and dual-layer freeform meta-atoms by stacking the top and bottom single-layer meta-atoms. (c)-(f) Zoomed-in SEM images showing the distribution of meta-atoms of three types and details of freeform meta-atoms.

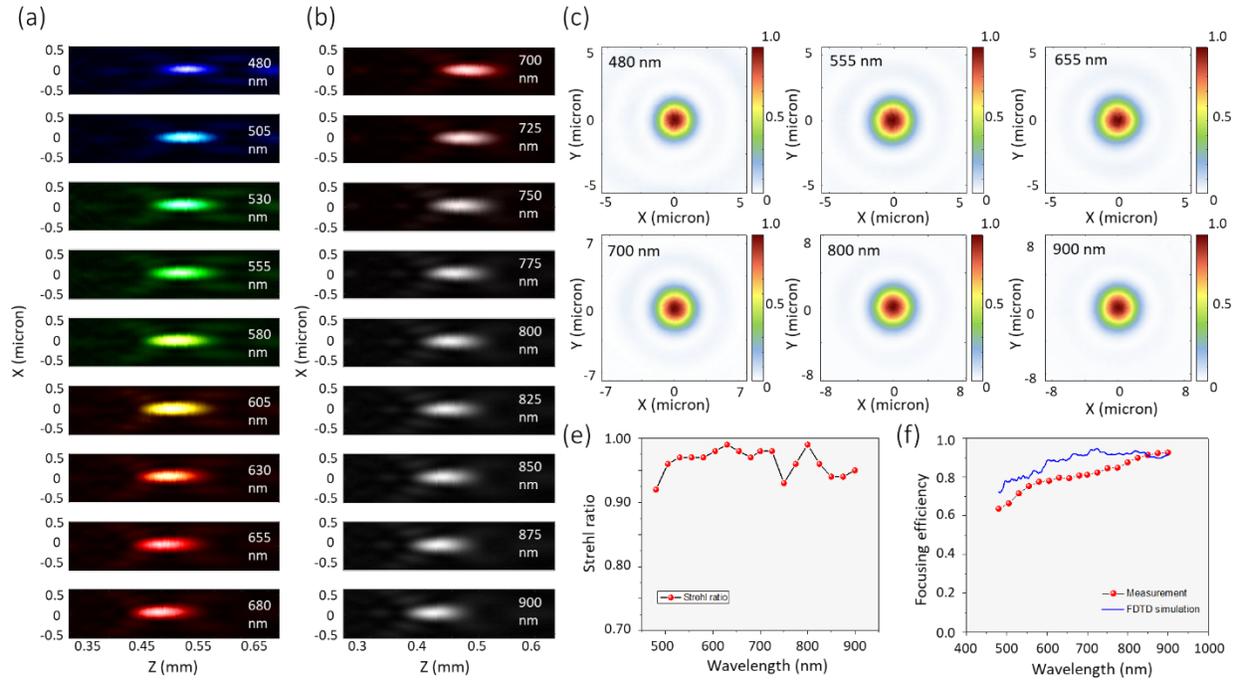

Figure 7. Measurement of the high-efficiency dispersion-engineered heterogeneous metalens. (a) Measured point spread functions in the XZ plane at wavelengths ranging from 480 nm to 680 nm. (b) Measured point spread functions in the NIR band from 700 nm to 900 nm in wavelength. (c) Measured normalized light intensity distributions in the focal planes at wavelengths of 480 nm, 555 nm, 655 nm, 700 nm, 800 nm, and 900 nm, respectively. (e) Measured Strehl ratios of focal spots in the VIS-NIR band. (f) Measured focusing efficiency of the dispersion-engineered metalens and the simulation results for comparison.